# Osmotically Driven Nonmonotonic Dynamics of Nuclear-to-Cellular Volume Ratio

Jize Sui*

State Key Laboratory of Nonlinear Mechanics, Institute of Mechanics, Chinese Academy of Sciences, Beijing 100190, China

## Abstract

A biophysical issue how the nuclear size dynamically scales with the cellular size remains mysterious. We develop a theoretical framework in which the interactions between polydisperse biomolecules and the mechanical elasticity of the cell are precisely integrated to investigate dynamics of the nuclear-to-cellular volume ratio (N/C ratio). We surprisingly find that the N/C ratio varies nonmonotonically with time, instead of maintaining a constant as normally known, during a period of osmotic shocks. Combining simulations and analytical argument, we identify that this nontrivial dynamics can be phenomenally predicted by the formed segregation configurations in the concentrations of the biomolecules, and essentially, all these observed phenomena can be further rationalized by the resultant of the excluded volume interactions between the polydisperse biomolecules and the spatial constraint from the nuclear envelope upon the macromolecule diffusions. Our results agree well with the published experiments for the cellular and nuclear size controls of the protoplasts.

Cells are the fundamental units that make up living organisms, especially the emergence of eukaryotes evolving more complex intracellular organizations and transport of living matters has made the kingdom of life flourishing [1,2]. The long-standing puzzles for the eukaryotic cell types are those whether the nuclear-to-cellular volume ratio, i.e., N/C ratio, always dynamically remains constant, as commonly known, during a cell cycle [3-5], and how to theoretically identify the physical mechanism responsible for the maintenance of the N/C ratio under the existing biological complexity that the cellular and the nuclear size control could highly depend on the self-integration and exchange of diverse biomolecules between the organelles inside the cell [3,6-8]. Recent studies suggested that the cell consisting of the nucleus behaves as an ideal nested osmometer due to their respective soft nature, and then the osmotic pressures created by the intracellular biomolecules, also termed as (bio)osmolytes, serves as a possible mechanism responsible for size controls of the nucleus and the cell [4,5,9]. However, previous models [4,5], based upon the Van't Hoff's relationship (volumetric scaling approach) at the osmotic equilibria state, have largely overlooked at least three factors which are of pivotal importance for most physiological scenarios. (i) The cells, as well as the nuclei, with cytoskeletons inside (inherent solid structures) are typically soft and elastic in responding to external osmotic shocks [10-12]; (ii) The biomolecules in the cells are polydisperse in size, in particular, macromolecular crowding has been frequently observed in both cytoplasm and nucleoplasm, highlighting the role of interactions between various biomolecules [7,13-17]; (iii) The processes of physiological activities in the cells usually proceed dynamically, representing the out-of-equilibrium nature of living systems [18].

In this Letter, we develop a new theoretical framework by fully considering these issues to study the dynamics of the N/C ratio in the cells exposed to the osmotic shocks. In our model, the cell and its nucleus are treated as the gel-like colloids, i.e., the microgel and the nanogel as depicted in Fig. 1, respectively. The crosslinked gel network (non-osmotic phase) represents the alike biofunctions of the cytoskeletons which ensure different elasticities for the cytoplasm and the nucleoplasm [2]. It showed previously that the excluded volume interactions between the polydisperse (bio)osmolytes hold vital importance in regulating their physiological stabilities [19,20] and biophase separations [21,22] in the intracellular crowding. Herein, we consider the intracellular plasma to be a binary mixture of small and large (bio)osmolytes, regarded as the hard colloids, and employ the depletion theory at thermodynamic equilibria in free energy density model for their interactions [16,23,24].

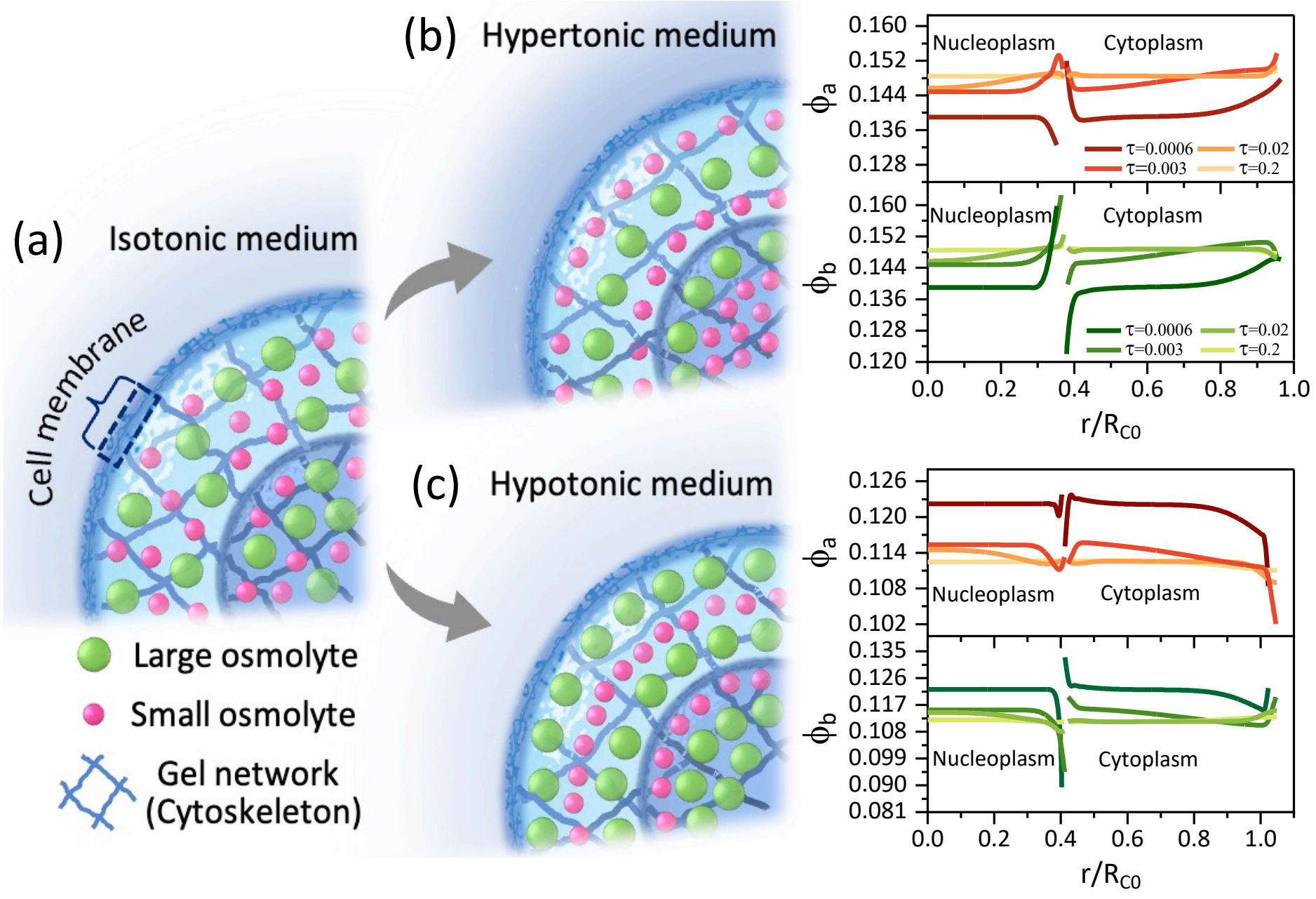

FIG. 1. Schematics of gel-like colloids mimicking the cell (microgel) consisted of the nucleus (nanogel). The cell under (a) isotonic condition, (b) hyperosmotic shifts and (c) hypoosmotic shifts. The concentration profiles in (b) and (c) manifest the respective segregated configurations of small and large osmolytes over time. Here, $\phi_{a0} = \phi_{b0} = 0.13$, $\phi_{w0} = 0.7$, $r_a = 3$ nm and $r_b = 20$ nm.

In this problem, when the cell experiences external osmotic shocks, only water can flow either inwards (for the hypoosmotic shifts) or outwards (for the hyperosmotic shifts) the cell across the cell membrane (CM), while small and large osmolytes are both restricted inside the cell. Within the cell, the nuclear envelope (NE), a semi-permeable membrane, allows the small osmolytes with a Stokes radius $r_a \leq 3$ nm, e.g., glycerol and nucleotide, to freely pass, while remains impermeable to the large osmolytes with the radius $r_b$ in the range of 6 nm to 40 nm, e.g., proteins, DNA and mRNA [4,5,15]. The flow of water not only alters the cellular volume, but also creates the coupling diffusive flows of the intracellular osmolytes. This enables the cell with an initial radius $R_{C0}$, as well as the nucleus with $R_{N0}$, to take their respective new sizes by $R_C(t)$ and $R_N(t)$. The kinematic equations of water, small and large osmolytes are governed by the force balance between the thermodynamic force, given by the gradient of the chemical potential, and the dissipative force, accounting for the hydrodynamic drag. Here, the components of water, gel network, small and large osmolytes are denoted by the subscripts: 'w', 'g', 'a' and 'b', respectively, and their individual volumes are $\sigma_i = 4\pi {r_i}^3/3$ with $r_i$ being the radii ($r_g$ means gyration radius of a segment in the gel network). The time-dependent volume fractions of these components are $\phi_i(r,t)$, and $\sum_i \phi_i = 1$. Since the components are same in the cytoplasm and the nucleoplasm, their kinematic equations can be represented universally within the bulk cell domain

$$\zeta_{wa}(V_w - V_a) + \zeta_{wb}(V_w - V_b) + \zeta_{wg}V_w + \nabla\mu_w = 0, \tag{1a}$$

$$\zeta_{aw}(V_a - V_w) + \zeta_{ab}(V_a - V_b) + \zeta_{ag}V_a + \nabla\mu_a = 0, \tag{1b}$$

$$\zeta_{bw}(V_b - V_w) + \zeta_{ba}(V_b - V_a) + \zeta_{bg}V_b + \nabla\mu_b = 0, \tag{1c}$$

where velocities (relative to the gel-network) are used as $V_i = v_i - v_g$ $(i = w, a, b)$ with $v_i$ being the components velocities relative to lab., and $\mu_i = \sigma_i\, \partial f/\partial\phi_i$ are the respective chemical potentials with $f$ being the free energy density model. In Eq. (1), due to the (semi)dilute regime (even for the intracellular crowding), the frictional coefficient for water molecules percolating through the osmolytes are much less than that for water molecules penetrating through the gel network $\zeta_{wg}$, i.e., $\zeta_{wa}$ and $\zeta_{wb}$ are ignored; the frictional coefficients of the osmolytes diffusing in water, the plasma and the gel network are all given by using Einstein relation $\zeta_{ij} = k_BT/D_{ij}$ $(i = a, b,\ j = w, a, b, g,\ i \neq j)$ with $D_{ij}$ being the self-diffusivities of the osmolytes. Herein, the osmolytes diffusivities in water are given with Stokes-Einstein-type $D_{iw} = k_BT/6\pi\eta r_i$, where $\eta$ is the water viscosity, and those in the gel network $D_{ig}$, as well as the $\zeta_{wg}$, are all fully discussed in section I in Ref. [25]. Determinations of the small and the large osmolytes diffusivities in the intracellular plasma, i.e., $D_{ab}$ and $D_{ba}$, however, are not easy. The microrheological experiments in vivo by using genetically encoded multimeric (GEM) nanoparticles clarified that the diffusivity of a certain osmolyte in the plasma often exponentially depends on the concentrations of other types of osmolytes, as fitted well by Phillies' model [5]. In this work, we therefore straightforward dictate these two diffusivities by

$$D_{ab} = D_{aw}\, exp(-k\beta_a\phi_b), \tag{2a}$$

$$D_{ba} = D_{bw}\, exp(-k\beta_b\phi_a), \tag{2b}$$

where the scaling parameters are suggested as $\beta_i = r_i\beta_{40}/20$ ($i = a, b$) with $\beta_{40} \approx 7.4$ measured for the GEMs with a diameter 40 nm (a mesoscale length for many biological entities) [15]. Note that, these diffusivities in Eq. (2) are often measured on the magnitude of $10^{-13}\ m^2/s$ ($D_{iw} \simeq 10^{-11}\ m^2/s$ for $r_i \leq$25 nm) [5,15], which can be matched by taking factor $k = 10$ in our model. The free energy density model $f(\phi_w, \phi_a, \phi_b, \phi_g)$ can be simply given akin to that for the colloidal gel solutions [23,26]

$$f = k_BT\left(\sum_{i=w,a,b}\frac{1}{\sigma_i}\phi_i\, ln\,\phi_i + \sum_{\substack{i=a,b,\\ j=a,b}}\frac{1}{\sigma_i\sigma_j}a_{ij}\phi_i\phi_j + G_0\left(\frac{\phi_g-\phi_{g0}}{\phi_{g0}}\right)^2\right), \tag{3}$$

where $a_{ij} = (2\pi/3)(r_i + r_j)^3$ is the second-order virial coefficient to represent the excluded volume interactions between the osmolytes, and the third term accounts for the elastic energy due to the small deformation of gel-like cytoskeletons with the initial concentration $\phi_{g0}$ in the cell, where $G_0 = mk_BT/\sigma_g$ is the elastic modulus with $m$ crosslinking degree.

Solving Eq. (1) by substituting Eqs. (2) and (3) and the bulk volumetric flux condition $\sum_{i=w,a,b,g} v_i\phi_i = 0$ inside the cell, we obtain the velocities $V_i$ for water and the osmolytes (for their complex expressions see Ref. [25]). One then arrives at the time evolution equations of $\phi_i$ inside the cell with 1-D spherical symmetry

$$\dot{\phi}_i = -\frac{1}{r^2}\frac{\partial(r^2J_i)}{\partial r}, \tag{4}$$

where $J_i = V_i\phi_i$ are the fluxes. Here, the use of velocities $V_i$ instead of the velocities $v_i$ is to implement "Soft-Cell" approach (SCA) for solving Eq. (4) numerically with the appropriate boundary conditions. The SCA we proposed previously displays an advantage of circumventing difficulties in addressing moving boundary complexes for gel dynamics [24,27]. The boundary conditions used here have become concise: (i) At the nucleus center $r = 0$, no fluxes exist $J_i = 0$; (ii) At the CM $r = R_C$, the fluxes of the osmolytes are zero $J_a = J_b = 0$, whereas water flux is related to the cell radius by $J_w = -\dot{R}_C(t)$; (iii) An extra condition is needed at the NE as $J_b|_{r=R_N} = 0$; (iv) The concentration gradient of the large osmolytes could become very large at the NE (a bio-interface) due to their cutoff diffusions, and this singular point shall cause the unexpected instability for the continuous diffusion equations of the small osmolytes and water across the NE. Then, for simplicity, we assume $\partial\phi_b/\partial r = 0$ only at the NE boundary in their diffusion equations. In our model, the NE has been assumed to be infinitely thin such that its interfacial energy is negligible [4,5], while the CM thickness varies with the cell size by $h = h_0(R_C/R_{C0})^{-\alpha}$ with $h_0$ an initial CM thickness [11]. The cellular size control, as validated experimentally [5,11,9], exclusively relies on the osmotic pressure difference causing the water flux passing through the CM, which can be determined by Darcy's law $\dot{R}_C(t) = \kappa\Delta\Pi/\eta h$, where $\kappa$ is the CM permeability, $\Delta\Pi = \Pi_{cyt} - \Pi_{sur}$ is the osmotic pressure difference between the cytoplasm and the surroundings. Strictly, $\Pi_{cyt}(t)$ is time-dependent since it is determined by the concentrations of solid phases $\phi_{a,b,g}(t)$ in the cytoplasm, leading to the difficulty in formulating $R_C$. If we first crudely assume $\Delta\Pi$ a constant, the cell radius can be expressed analytically for $\alpha = 1$ by

$$R_C/R_{C0} = exp(t\Delta\Pi\kappa/\eta h_0 R_{C0}), \tag{5}$$

which typifies an exponential mode of cellular size control, as normally identified in the experiments [9,15]. For the sake of simplicity, we pre-apply this given exponential form of $R_C$ for the boundary condition, while during the SCA calculations, we revisit the time-dependent nature of $\Pi_{cyt}$ in Eq. (5) by

$$\Pi_{cyt}(t) = nk_BT \sum_{i=a,b} \frac{\bar{\phi}_i}{\sigma_i} z(\bar{\phi}_i), \tag{6}$$

where the contribution of the cytoskeletons in the pressure is ignored due to its very dilute concentration (e.g., $\phi_{g0}$ =0.04), n is the matching factor, and the solution compressibility factors are given by $z(\bar{\phi}_i) = \frac{1+a_1\bar{\phi}_i+a_2\bar{\phi}_i^2+a_3\bar{\phi}_i^3}{1-\bar{\phi}_i/\phi_p}$, where the coefficients are $a_1 = 4 - 1/\phi_p$, $a_2 = 10 - 4/\phi_p$ and $a_3 = 18 - 10/\phi_p$, and $\phi_p$ =0.58 being the loose packing fraction [26,28]. The osmolytes concentrations used in Eq. (6) are volumetric average in the cytoplasm $\bar{\phi}_i(t) = 3\int_{R_N}^{R_C} r^2\phi_i dr/(R_C^3 - R_N^3)$. A set of coupled diffusion equations in Eq. (4) can be made dimensionless with the length scaled by $R_{C0}$ and time by $\tau = t\,D_{aw}/R_{C0}^2$ for SCA calculations (for the intact procedures see section III in Ref. [25]). The dimensionless cellular size is then $R_C(\tau)/R_{C0} = exp(Cont\,\Delta\tilde{\Pi}\,\tau)$ with $\Delta\tilde{\Pi} = \sigma_a\Delta\Pi/k_BT$ and $Cont = 5\kappa R_{C0}/h_0\,r_a^2$. We dictate some main parameters fixed in this problem: $R_{C0}$=3 $\mu m$, $h_0$=100 nm, $r_w$ =0.2 nm, $r_a$ =3 nm, $r_g$ =30 nm, $\kappa \approx 7\times10^{-19}$ m$^2$ and $R_{N0}/R_{C0}$ =0.406. The intracellular elastic modulus is complex depending on cell types, but the nucleus may be normally softer than the cell [2,12]. We then subdivide $G_0$ into $G_0^N \approx$0.11 kPa in the nucleoplasm and $G_0^C \approx$10 kPa in the cytoplasm. We set water initial concentration $\phi_{w0} = 0.7$, and the equal initial concentrations of the osmolytes $\phi_{a0} = \phi_{b0}$ =0.13. This produces, revisiting Eq. 6, an initial isotonic pressure in the cytoplasm $\Pi^0_{cyt}$ ≈0.923 MPa for $r_b$ =20 nm ($n$=110) [15].

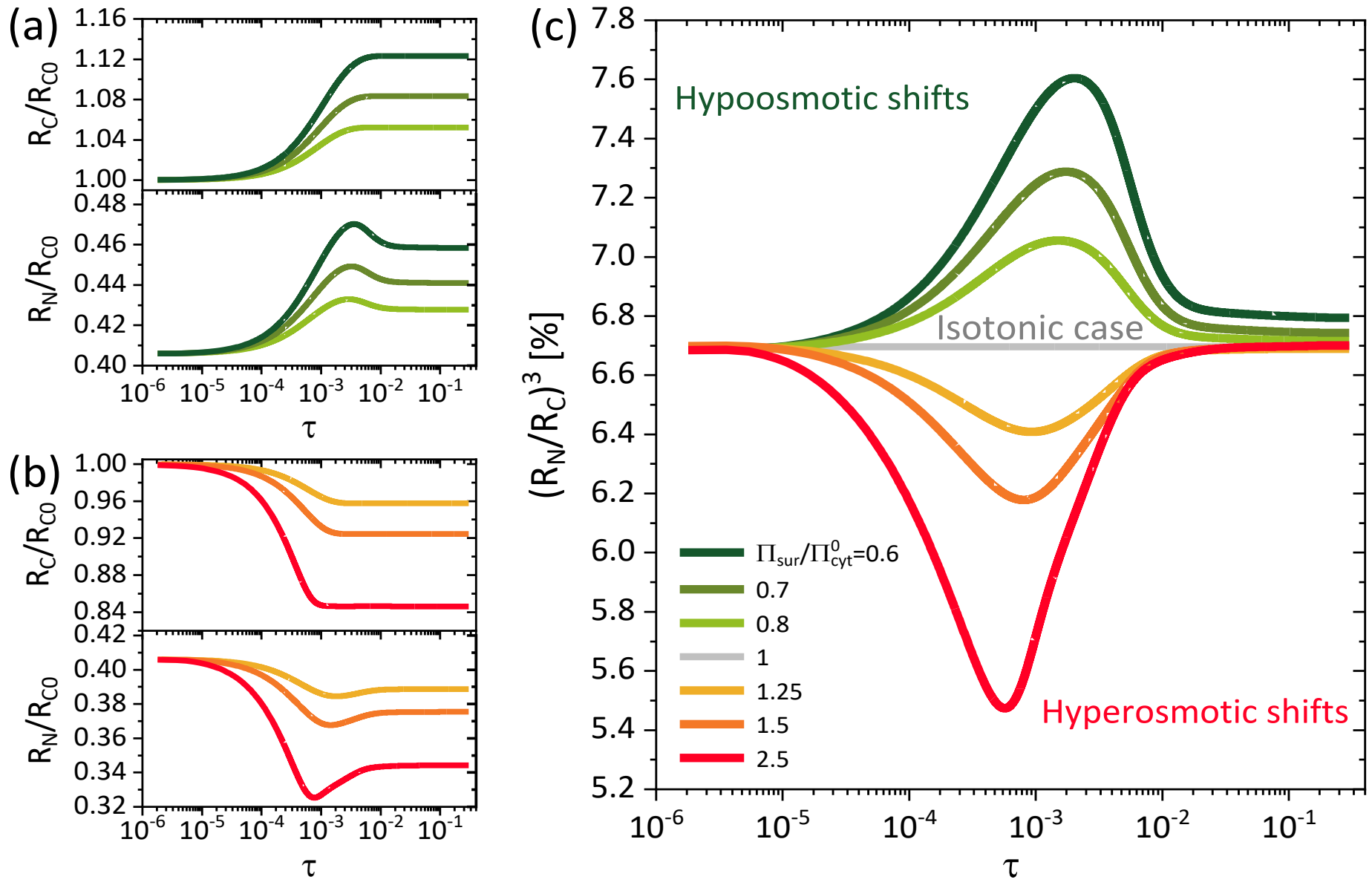

FIG. 2. Dynamics of the cellular size control and the nuclear size control for (a) hypoosmotic shifts and (b) hyperosmotic shifts. (c) Non-monotonic dynamics of the N/C ratio under variable osmotic shifts. The initial N/C ratio $\varepsilon_0$ = 6.69% and $r_b$ =20 nm.

We first examine the influences of the osmotic shocks. As Figs. 1(b) and 1(c) shown, the hypertonic shock raises the osmolytes concentrations, while the hypotonic shock lowers them in the cell. We found that the hyperosmotic shocks, leading to shrinkage of the cell [Fig. 1(b)], can create the concentration pattern in the cytoplasm where the small osmolytes enrich (monotonic increases in $\phi_a$) at the domains near the CM and the NE, while the large osmolytes are driven towards the middle domain, and meanwhile, the pattern in the nucleoplasm where the large osmolytes enrich (monotonic increases in $\phi_b$) towards the NE, while the small osmolytes are repelled away from the NE (a negative gradient of $\phi_a$). These pattern configurations under the hypoosmotic shocks by which the cell swells [Fig. 1(c)], however, are seen to behave conversely in the domains corresponding to those under the hyperosmotic shocks. These resulting patterns, indicative of the osmolytes segregation configurations in space, form in accompanying with noticeable concentration differences of the osmolytes near the NE at the earlier stage, whereases they vanish gradually as the cell arrives at the new steady state by adapting to the osmotic shocks, when the osmolytes concentrations become comparable again inside and outside the nucleus. The self-interactions (virial coefficients $a_{aa}$ and $a_{bb}$ in Eq. (3)) and the cross-interactions (virial coefficients $a_{ab} = a_{ba}$ in Eq. (3)) between the osmolytes both determine their motions [25]. The strong size asymmetry of the osmolytes enhances the cross-interactions whose effect on the osmolytes diffusions is not symmetric, which essentially leads to the formation of specific segregation configurations. The exhaustive analyses, which are also available for our problem, can be cited in the work on the stratification in drying the binary colloidal mixtures [29].

Figure 2 shows the dynamics of cellular size, the nuclear size and the N/C ratio for different osmotic shocks. Generally, the hyperosmotic shocks cause both the cell and nucleus to shrink until to the osmotic balance, and vice versa. Comparing to the $R_C$, the $R_N$ displays the nonmonotonic variations over time, e.g., an increase followed by a slight decrease under the hypoosmotic shocks [Fig. 2(a)], or a decrease followed by a slight increase under the hyperosmotic shocks [Fig. 2(b)]. We further surprisingly find that the N/C ratio varies nonmonotonically over time, instead of a robustly maintained constant, during the osmotic treatment period [Fig. 2(c)]. For the hyperosmotic shocks, the N/C ratio first decreases to a minimum and then increases again, while it behaves conversely for the hypoosmotic shocks. Especially, under the hyperosmotic shocks (or slight hypoosmotic shocks), the N/C ratio at the steady state can take the same value as the initial one. Although formulating an analytical expression finely fitting the numerical profiles of N/C ratio in Fig. 2(c) has remained elusive (involving the intrinsically coupled diffusions of intracellular matters), we can still analytically rationalize the universal mechanism responsible for this nontrivial dynamics of the N/C ratio. Let $\varepsilon(t) = {R_N}^3/{R_C}^3$ be the N/C ratio. Since the large osmolytes in the nucleoplasm and the cytoplasm are conserved in numbers, we have ${R_N}^3 = {R_{C0}}^3\,\phi_{b0}^{nuc}/\bar{\phi}_b^{nuc}$ and ${R_C}^3 - {R_N}^3 = \left({R_{C0}}^3 - {R_{N0}}^3\right)\phi_{b0}^{cyt}/\bar{\phi}_b^{cyt}$, where $\bar{\phi}_b^{nuc}(t)$ and $\bar{\phi}_b^{cyt}(t)$ are assumed to be the volumetric average concentrations with spatial independence. The N/C ratio is then written as $\varepsilon(t) = \varepsilon_0/(\varepsilon_0(1-k)+k)$, where $\varepsilon_0 = R_{N0}/R_{C0}$ is an initial N/C ratio and $k = \phi_{b0}^{cyt}\bar{\phi}_b^{nuc}/\phi_{b0}^{nuc}\bar{\phi}_b^{cyt}$. Our calculations confirm that the factor $k$ can closely approximate to 1 for a wide range of concentrations considered here [25]. Therefore, the N/C ratio can be estimated by $\varepsilon(t) \cong \varepsilon_0/k$. As mentioned earlier, the calculations in Fig. 2(c) were performed by assuming the equal initial concentrations $\phi_{b0}^{nuc} = \phi_{b0}^{cyt}$, also denoting the comparable initial concentrations of the non-osmotic phases (gels). With this special condition, the concentrations $\bar{\phi}_b^{nuc}$ and $\bar{\phi}_b^{cyt}$ are seen to be finally comparable at the steady state [Figs. 1(b) and 1(c)], i.e., $k = 1$, which leads to $\varepsilon(t \to \infty) = \varepsilon_0$. Therefore, if the large osmolytes localized to the nucleoplasm and the cytoplasm are initially equal in concentrations, their concentrations most likely maintain to be comparable as the cell takes the new steady state (also see Fig. S3 in Ref. [25]). This regime ($k = 1$) naturally allows the N/C ratio at the steady state to maintain the same value as its initial one. Our results show that this N/C ratio maintenance is more robust for the cells exposed to the hyperosmotic shocks (similar to the macromolecular crowding) than the hypoosmotic shocks [Fig. 2(c)].

Besides, the N/C ratio maintenance is shown to remain unaffected by the factors such as the osmotic shocks [Fig. 2(c)], the cellular elasticity [Fig. 3(c)], though these factors significantly alter the N/C ratio dynamics. In reality, the similar concentrations of the large osmolytes (or non-osmotic phases) in between the nucleoplasm and the cytoplasm, while only a special case in our treatment, have been frequently suggested for many cell types [4,5]. For a more general case $\phi_{b0}^{nuc} \neq \phi_{b0}^{cyt}$ which also indicates $\bar{\phi}_b^{nuc} \neq \bar{\phi}_b^{cyt}$ at the steady state, we find the visible deviations of $\varepsilon(t \to \infty)$ from $\varepsilon_0$ (Fig. S4 (a) in Ref. [25]). We shall suggest here another form to readily evaluate $\varepsilon(t)$ for this general case

$$\varepsilon(t) \cong \varepsilon_0 \frac{\phi_{b0}^{cyt}}{\phi_{a0}} \frac{\bar{\phi}_a(t)}{\bar{\phi}_b^{cyt}(t)}, \tag{7}$$

where we have adopted the conservation law of the small osmolytes within the cell ${R_C}^3 = {R_{C0}}^3\, \phi_{a0}/\bar{\phi}_a(t)$. Eq. (7) manifests an experimentally feasible avenue in which the N/C ratio, especially at the steady state, can be estimated well by only accessing the information, such as the average concentrations of the osmolytes, in the cytoplasm instead of the nucleoplasm.

To further explore the physical origin of the observed nonmonotonic dynamics of the N/C ratio, we start from the time derivative of the N/C ratio $\dot{\varepsilon} = 3\varepsilon\big(\dot{R}_N/R_N - \dot{R}_C/R_C\big)$, and naturally $\dot{\varepsilon} < 0$ for a decreasing $\varepsilon$, while $\dot{\varepsilon} > 0$ for an increasing $\varepsilon$. The Sgn($\dot{\varepsilon}$) (the sign of $\dot{\varepsilon}$) merely depends on the Sgn($\dot{R}_N/R_N - \dot{R}_C/R_C$). By revisiting the conservation equations in Eq. (4), if considering the concentrations independent of $r$, we can rewrite them as $-J_i(t,r) = r\dot{\phi}_i/3$ ($i = a, w$). The cellular size and the nuclear size are related to the fluxes of the small osmolytes and water by $\dot{R}_C = -J_w(t, R_C)$ and $\dot{R}_N = -\big(J_a(t, R_N) + J_w(t, R_N)\big)$, respectively. We then obtain $\dot{R}_C/R_C \cong \dot{\phi}_w$ and $\dot{R}_N/R_N \cong \dot{\phi}_a + \dot{\phi}_w$, and therefore arrive at $\dot{R}_N/R_N - \dot{R}_C/R_C \cong \dot{\phi}_a$. This produces

$$Sgn(\dot{\varepsilon}(t)) \Leftrightarrow Sgn\big(-J_a(t, R_N)\big). \tag{8}$$

It is clear that the flux of the small osmolytes across the NE can phenomenally predict the nonmonotonic behaviors of the N/C ratio. Figures 3(a) and 3(b) show the flux of the small osmolytes across the NE varying with time for the hyperosmotic and the hypoosmotic shocks, respectively. Clearly, the hyperosmotic shocks lead to the efflux of the small osmolytes, as well as water, from the nucleus, i.e., $J_a(R_N) > 0$, at the earlier stage, while as the specific segregations of the osmolytes are formed near the NE [Fig. 1(b)], in which the concentration of the small osmolytes at the outer side of NE becomes higher than that at the inner side of NE, the small osmolytes will diffuse backward the nucleus, i.e., an influx towards the nucleus $J_a(R_N) <$ 0, at the later stage. This elucidates that, by revisiting Eq. (8), $Sgn(\dot{\varepsilon}) < 0$ is followed by $Sgn(\dot{\varepsilon}) > 0$ over time for the hyperosmotic shocks, indicating the downward convex profiles of the N/C ratio [Fig. 2(c)]. The similar analysis [Fig. 3(b)] could be applied to understand the upward convex profiles of the N/C ratio for hypoosmotic shocks [Fig. 2(c)]. With these results, we argue that the excluded volume interactions between the intracellular osmolytes of different sizes and also the preferential localization of the large osmolytes constrained by the NE mutually determine transitions between the efflux or the influx of the small osmolytes from the nucleus, which essentially manage the nonmonotonic configurations of the N/C ratio dynamics under the osmotic shocks.

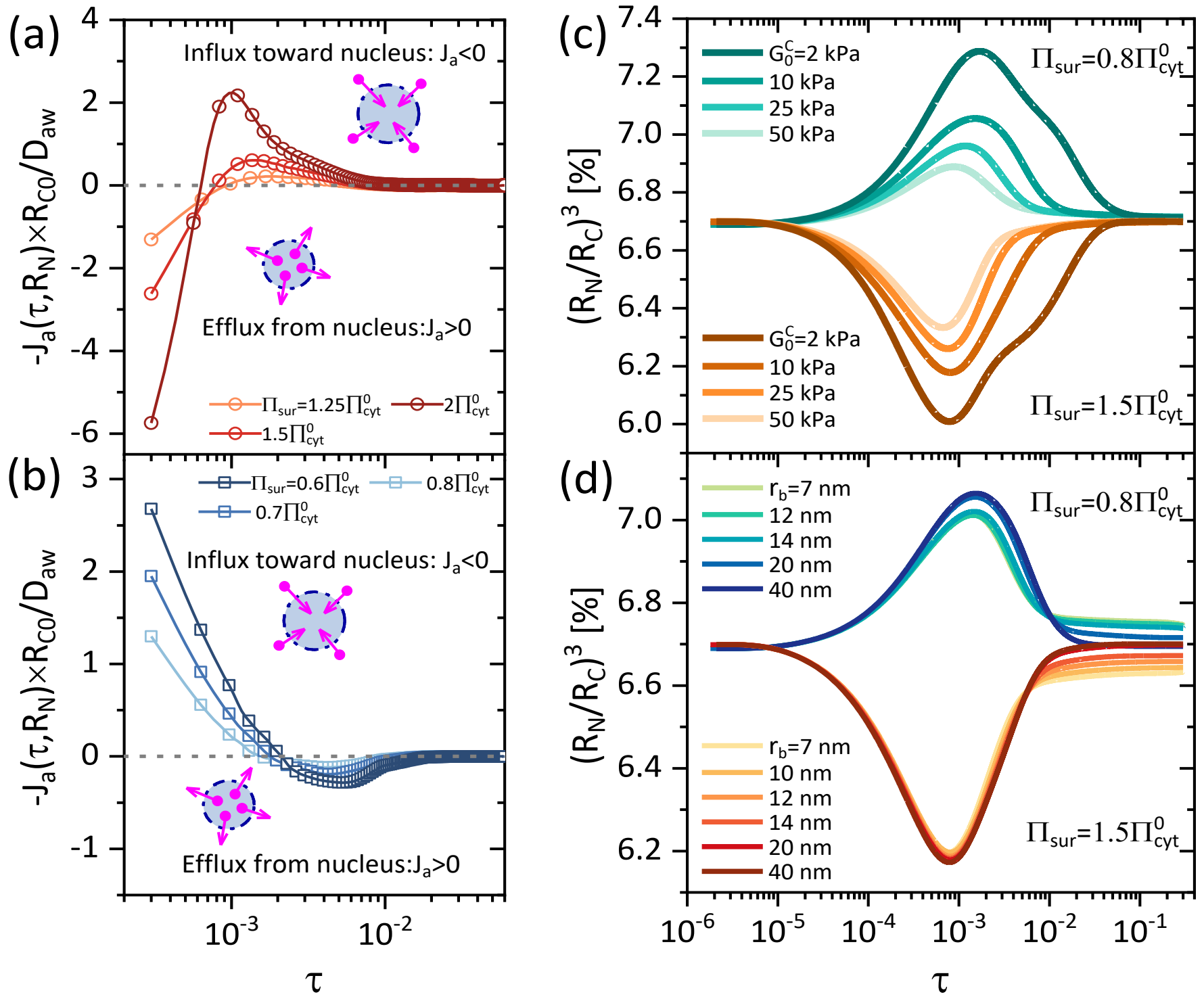

FIG. 3. Volumetric flux of the small osmolytes across the NE $J_a(\tau, R_N)$ (scaled by $R_{C0}/D_{aw}$) varies with time under (a) hyperosmotic shocks and (b) hypoosmotic shocks. The used parameters are the same as in Figs. 1 and 2. The N/C ratio dynamics can be modulated by (c) the cellular elasticity and (d) the radius of large osmolyte under the osmotic shocks.

We also probe the effects of the cytoplasm elasticity $G_0^C$ (roughly regarded as the cellular elasticity) and the radius of the large osmolyte on the N/C ratio dynamics in Figs. 3(c) and 3(d), respectively. As the cell gets softer, the N/C ratio can experience the smaller minimum (for the hyperosmotic shifts) or the larger maximum (for the hypoosmotic shifts) over time, while its value at the steady state remains unaffected. Instead, the large osmolyte radius can markedly alter the N/C ratio at the steady state rather than its dynamics. It is indicated that the biological factors, such as the concentration and the size of the osmolytes, which are all involved in the depletion interactions between the osmolytes could break the maintenance of the N/C ratio at the steady state. Figures 4(a) and 4(b) further show that the extremums of the N/C ratio $\varepsilon_{ext}$, as well as its corresponding normalized time $\tau_{ext}$ at which the N/C ratio change its monotonicity, are mainly modulated by the osmotic pressure and the cellular elasticity, which both follow well the power law fittings. We also note that the time $\tau_{ext}$ of the N/C ratio is earlier than that of the nuclear size control [Fig. 4(a)] for all the osmotic shocks, indicating that the nonmonotonic behaviors of the N/C ratio do not arise from the nonmonotonic variations in nuclear radius.

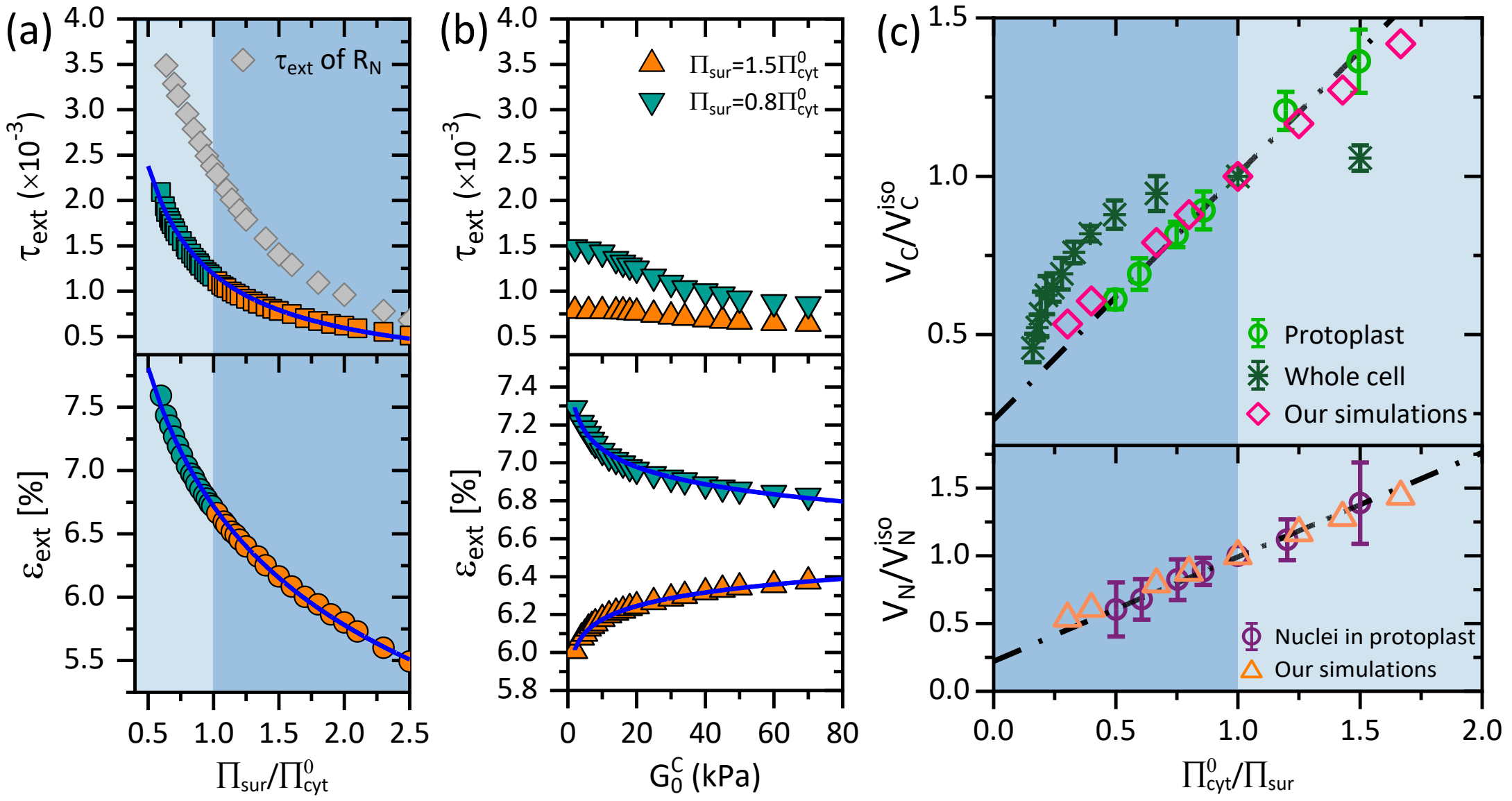

FIG. 4. Extremums of the N/C ratio $\varepsilon_{ext}$ and the corresponding normalized time $\tau_{ext} = ct_{ext}k\, D_{aw}/{R_{C0}}^2$ both vary as the functions of (a) the external osmotic pressure and (b) the cellular elasticity ($\tau_{ext}$ is affected little by $G_0^C$). The parameters are the same as in Fig.2. The blue solid curves are best-fit power laws in (a): $y = 0.001x^{-1}$ for $\tau_{ext}$ and $y = 0.067x^{-0.22}$ for $\varepsilon_{ext}$ and in (b): $y = 0.06x^{0.016}$ for the hyperosmotic shock and $y = 0.07x^{-0.018}$ for the hypoosmotic shock. All fitting errors are about $\pm 2 \times 10^{-4}$. (c) Comparisons between the experiments (symbols with error bars) and our simulations for the cellular volume $V_C/V_C^{iso}$ and the nuclear volume $V_N/V_N^{iso}$, where $V_C^{iso}$ and $V_N^{iso}$ are the volumes in the isotonic medium, as the functions of the osmotic shocks. The experiments are taken from Ref. [5], and the dashed lines, i.e., Boyle Van't Hoff (BVH) fitting, depict the behaviors of the ideal osmometers in absence of membrane tension from the elastic cell wall [5]. The initial N/C ratio $\varepsilon_0 = 8\%$, and other parameters are the same as in Fig.1.

At last, Fig. 4(c) illustrates the comparisons between our model and the osmotic shock experiments for the cellular and the nuclear volumes measured from the whole fission yeast cells and the protoplasts [5]. The whole cell with the elastic cell wall exerting interfacial stress behaves an apparently nonlinear dependence of the volume on the osmotic shocks, whereas the protoplast, the cell whose cell wall has been removed, changes its volume, as well as its nuclear volume, linearly with the osmotic shocks. Our model, which has been established by ignoring the elastic interfacial energy from the cell wall for simplicity, is near identical to the case of protoplasts. In fact, though we simply refer to the exponential mode of cellular size control in Eq. (5), the employment of the time-dependent osmotic pressure $\Pi_{cyt}(t)$ in Eq. (6) prevents us from formulating the cellular size analytically, suggesting that the cellular size at the state of the osmotic shock adaption can be finally accessed by the numerical measurements. The results in Fig. 4(c) that our numerical measurements excellently agree with the linear law validated by the experiments have testified the proposed model a robust paradigm in addressing the cellular and the nuclear size controls driven by the osmotic shocks.

In conclusion, we have presented a theoretical framework to investigate the dynamics of nuclear-to-cellular volume ratio. Our results primarily confirm the surprising finding that the N/C ratio can vary nonmonotonically over time under the osmotic shocks. Phenomenologically, this nontrivial dynamics of the N/C ratio can be predicted by the transition between the efflux and the influx of the small osmolytes from the nucleus, which is dominated by the specific segregation configurations formed by the polydisperse osmolytes in space ranging from the nucleoplasm to the cytoplasm. The biophysical origin of these observed phenomena is essentially attributed to the resultant of the excluded volume interactions between the osmotically active osmolytes and the preferential localizations of the macromolecules separated by the NE. Our findings can be supported by the equilibrium theories and the protoplasts experiments [4,5]. For example, when the large osmolytes (or non-osmotic phase) are initially comparable in concentrations in between the cytoplasm and the nucleoplasm, the N/C ratio maintains the value at the steady state the same as the initial one, which remains unaffected by the osmotic shocks and cellular elasticity. It would be intriguing to extend our simple framework for integrating more open issues, such as homeostasis regimes [5,15] and bio-interfacial dynamics [30,31]. We hope the essential principles presented here will advance our understandings of biology-related dynamic processes, e.g., the N/C ratio, macromolecular crowding, bio-phase separations, which hold physiological importance for the living organisms.

This work acknowledges the support from the National Natural Science Foundation for Young Scientists of China, Grant No. 21903003. The author declares no competing financial interest.

[1] Toni Gabaldón, Annu. Rev. Microbiol. **75**, 631 (2021).
[2] M. E. Allen, J. W. Hindley, D. K. Baxani, O. Ces, and. Y. Elani, Nat. Rev. Chem. **6**, 562 (2022).

[3] Y. Kalukula, A. D. Stephens, J. Lammerding, and S. Gabriele, Nat. Rev. Mol. Cell Biol. **23**, 583 (2022).
[4] D. Deviri, and S. A. Safran, Proc. Natl. Acad. Sci. U.S.A. **119**, E2118301119 (2022).
[5] J. Lemière, P. Real-Calderon, L. J. Holt, T. G. Fai, and F. Chang, eLife **11**, E76075 (2022).
[6] A. S. Verkman, Trends Biochem. Sci. **27**, 27 (2002).
[7] G. Rivas, and A. P. Minton, Trends Biochem. Sci. **41**, 970 (2016).
[8] J. Berry, C. P. Brangwynne, and M. Haataja, Rep. Prog. Phys. **81**, 046601 (2018).
[9] C. Cadart, L. Venkova, P. Recho, M. C. Lagomarsino, and M. Piel, Nat. Phys. **15**, 993 (2019).
[10] D. A. Fletcher, and R. D. Mullins, Nature **463**, 485. (2010).
[11] H. Li, Y. Zheng, Y. L. Han, S. Q. Cai, and M. Guo, Proc. Natl. Acad. Sci. U.S.A. **118**, E2022422118 (2021).
[12] S. Grosser, J. Lippoldt, L. Oswald, M. Merkel, D. M. Sussman, F. Renner, P. Gottheil, E. W. Morawetz, T. Fuhs, X. F. Xie, S. Pawlizak, A. W. Fritsch, B. Wolf, L.-C. Horn, S. Briest, B. Aktas, M. L. Manning, and J. A. Käs, Phys. Rev. X **11**, 011033 (2021).
[13] C. A. Brackley, M. E. Cates, and D. Marenduzzo, Phys. Rev. Lett. **111**, 108101 (2013).
[14] G. Nettesheim, I. Nabti, C. U. Murade, G. R. Jaffe, S. J. King, and G. T. Shubeita, Nat. Phys. **16**, 1144 (2020).
[15] B. Alric, C. Formosa-Dague, E. Dague, L. J. Holt, and M. Delarue, Nat. Phys. **18**, 411 (2022).
[16] A. Pastore, and P. A. Temussi, Trends Biochem. Sci. **47**, 1048 (2022).
[17] T. Skóra, M. Janssen, A. Carlson, and S. Kondrat, Phys. Rev. Lett. **130**, 258401 (2023).
[18] S. K. Gupta, and M. Guo, J. Mech. Phys. Solids **107**, 284 (2017).
[19] D. Gnutt, M. Gao, O. Brylski, M. Heyden, and S. Ebbinghaus, Angew. Chem. Int. Ed. **54**, 2548 (2015).
[20] G. Bajpai, D. A. Pavlov, D. Lorber, T. Volk, and S. Safran, eLife **10**, E63976 (2021).
[21] Y. Zhang, D. S. W. Lee, Y. Meir, C. P. Brangwynne, and N. S. Wingreen, Phys. Rev. Lett. **126**, 258102 (2021).
[22] J. D. Wurtz, and C. F. Lee, Phys. Rev. Lett. **120**, 078102 (2018).
[23] J.-P. Hansen, and I. R. McDonald, Theory of simple. liquids with applications to soft matter (Fourth Edition), Academic Press of Elsevier, Oxford, UK, 2013.
[24] J. Z. Sui, Phys. Chem. Chem. Phys. **25**, 410 (2023).
[25] See Supplemental Material.
[26] J. Z. Sui, Phys. Chem. Chem. Phys. **22**, 14340. (2020).
[27] J. Z. Sui, Phys. Rev. E **106**, 054701 (2022).
[28] J. Z. Sui, Phys. Rev. E **99**, 062606 (2019).
[29] J. J. Zhou, Y. Jiang, and M. Doi, Phys. Rev. Lett. **118**, 108002 (2017).
[30] L. M. C. Sagis, Rev. Mod. Phys. **83**, 1367 (2011).
[31] B. Gouveia, Y. Kim, J. W. Shaevitz, S. Petry, H. A. Stone, and C. P. Brangwynne, Nature **609**, 255 (2022).

# Supplemental Material

Osmotically Driven Nonmonotonic Dynamics of Nuclear-to-Cellular Volume Ratio

Jize Sui*

State Key Laboratory of Nonlinear Mechanics, Institute of Mechanics, Chinese Academy of Sciences, Beijing 100190, China

## I: Expressions of the velocities of water, small and large osmolytes

The friction coefficient of water molecules penetrating through the gel-like cytoskeletons is represented as $\zeta_{wg} = \phi_w \sigma_w \eta_w / \kappa_{wg}$, where $\kappa_{wg}$ is the permeability of the gel-network, $\sigma_w$ water molecule volume and $\eta_w$ water viscosity. The permeability is given here by adopting a common (semi)empirical form [1,2]

$$\kappa_{wg} = \frac{\xi^2}{K_g} \phi_w, \tag{S1}$$

where $\xi$ is the correlation length interpreted as the average mesh size in the gel-network and it can be expressed by a simple scaling law $\xi = b r_g {\phi_g}^{\beta}$ with $r_g$ being the gyration radius of a segment in the gel-network at the reference state, pre-factor is $b = 1$, the exponent takes $\beta = -0.75$ in a good solvent [3], and the fitting constant $K_g$ is a large value usually determined by the experimental measurements, and we assume it as $K_g = 10^3$ in our problem.

In the main text, we have considered the diffusivity of the small osmolyte with a radius $r_a$ =3 nm in water as a constant $D_{aw} = k_B T / 6\pi\eta_w r_a \approx 9.25 \times 10^{-11}\ m^2/s$ at a temperature 30 ℃. The diffusivity of the large osmolyte in water $D_{bw} = k_B T / 6\pi\eta_w r_b$ varies since its radius $r_b$ is variable. Using Einstein's relation, as discussed in the main text, the friction coefficients of the small and the large osmolytes penetrating through the gel-like cytoskeletons are given by $\zeta_{ag} = k_B T / D_{ag}$ and $\zeta_{bg} = k_B T / D_{bg}$, where $D_{ag}$ and $D_{bg}$ are the self-diffusivities of the small and the large osmolytes in the gel-network, respectively. These diffusivities $D_{ag}$ and $D_{bg}$ are generally associated with the percolated nature of crosslinked biopolymer network in the cells and can be complicated.

We address such a complexity in $D_{ag}$ and $D_{bg}$ by utilizing the empirical expression suggested by R. Hołyst et al. for the microgels (or nanogels) [3], yielding

$$\frac{D_{aw}}{D_{ag}} = exp\left(\frac{r_{eff}^a}{\xi}\right)^a, \qquad \frac{D_{bw}}{D_{bg}} = exp\left(\frac{r_{eff}^b}{\xi}\right)^a, \tag{S2}$$

where the fitting exponent $a$ could be assumed to be 1 (whose physical meaning is still under discussion [3]), and $r_{eff}^{a}$ and $r_{eff}^{b}$ are the effective hydrodynamic radii. In Eq. (S2), two effective hydrodynamic radii can be conducted by the correlation functions ${r_{eff}^{a}}^{-2} = {r_a}^{-2} + {r_g}^{-2}$ and ${r_{eff}^{b}}^{-2} = {r_b}^{-2} + {r_g}^{-2}$, yielding

$$r_{eff}^{a} = r_g\left(1+\left(\frac{r_g}{r_a}\right)^2\right)^{-\frac{1}{2}}, \qquad r_{eff}^{b} = r_g\left(1+\left(\frac{r_g}{r_b}\right)^2\right)^{-\frac{1}{2}}. \tag{S3}$$

Solving the kinematic equations for water, the small and the large osmolytes, i.e., Eqs. (1a)-(1c) in the main text, with above set of equations and the conditions $V_w\phi_w + V_a\phi_a + V_b\phi_b + v_g = 0$ and $\phi_w + \phi_a + \phi_b + \phi_g = 1$, we can obtain the velocities (relative to the gel-network) of water, the small and the large osmolytes, respectively

$$V_w = \frac{-5D_{aw}}{K_g{\lambda_{ag}}^2} d_w\left(df_{ww} * \frac{\partial\phi_w}{\partial r} + df_{wa} * \frac{\partial\phi_a}{\partial r} + df_{wb} * \frac{\partial\phi_b}{\partial r}\right), \tag{S4}$$

$$V_a = \frac{D_{aw}}{Q}\left[(d_{111}df_{ww} + d_{112}df_{wa} + d_{113}df_{wb}) * \frac{\partial\phi_w}{\partial r} + (d_{111}df_{wa} + d_{112}df_{aa} + d_{113}df_{ab}) * \frac{\partial\phi_a}{\partial r} + (d_{111}df_{wb} + d_{112}df_{ab} + d_{113}df_{bb}) * \frac{\partial\phi_b}{\partial r}\right], \tag{S5}$$

$$V_b = \frac{D_{aw}}{Q}\left[(d_{221}df_{ww} + d_{222}df_{wa} + d_{223}df_{wb}) * \frac{\partial\phi_w}{\partial r} + (d_{221}df_{wa} + d_{222}df_{aa} + d_{223}df_{ab}) * \frac{\partial\phi_a}{\partial r} + (d_{221}df_{wb} + d_{222}df_{ab} + d_{223}df_{bb}) * \frac{\partial\phi_b}{\partial r}\right], \tag{S6}$$

where all the pre-coefficients are represented as following

$$d_w = {\lambda_{aw}}^3(1-\phi_w-\phi_a-\phi_b)^{2\beta}, \tag{S7}$$

$$Q = -K_g{\lambda_{ab}}^2{\lambda_{ag}}^2\left(1 + e^{\beta_b\phi_a} + e^{\beta_a\phi_b} + FE1\left(1+e^{\beta_b\phi_a}\right) + FE2\left(1+e^{\beta_a\phi_b}\right) + FE1 * FE2\right), \tag{S8}$$

$$d_{111} = 5{\lambda_{aw}}^3{\lambda_{ab}}^2(1-\phi_w-\phi_a-\phi_b)^{2\beta}\left(1 + FE2 + e^{\beta_b\phi_a} + e^{\beta_a\phi_b}\right), \tag{S9}$$

$$d_{112} = K_g{\lambda_{ab}}^2{\lambda_{ag}}^2{\lambda_{aw}}^3\left(1 + FE2 + e^{\beta_b\phi_a}\right), \tag{S10}$$

$$d_{113} = K_g {\lambda_{ag}}^2 {\lambda_{aw}}^3 e^{\beta_a \phi_b}, \tag{S11}$$

$$d_{221} = 5{\lambda_{aw}}^3 {\lambda_{ab}}^2 (1 - \phi_w - \phi_a - \phi_b)^{2\beta} \left(1 + FE1 + e^{\beta_b \phi_a} + e^{\beta_a \phi_b}\right), \tag{S12}$$

$$d_{222} = K_g {\lambda_{ab}}^2 {\lambda_{ag}}^2 {\lambda_{aw}}^3 e^{\beta_b \phi_a}, \tag{S13}$$

$$d_{223} = K_g {\lambda_{ag}}^2 {\lambda_{aw}}^3 \left(1 + FE1 + e^{\beta_a \phi_b}\right), \tag{S14}$$

$$df_{ww} = \frac{\partial^2 f}{\partial {\phi_w}^2}, \quad df_{aa} = \frac{\partial^2 f}{\partial {\phi_a}^2}, \quad df_{bb} = \frac{\partial^2 f}{\partial {\phi_b}^2}, \tag{S15}$$

$$df_{wa} = \frac{\partial^2 f}{\partial \phi_w \partial \phi_a}, \quad df_{wb} = \frac{\partial^2 f}{\partial \phi_w \partial \phi_b}, \quad df_{ab} = \frac{\partial^2 f}{\partial \phi_a \partial \phi_b}. \tag{S16}$$

In above set of equations, the size ratios are $\lambda_{aw} = r_a/r_w$, $\lambda_{ab} = r_a/r_b$ and $\lambda_{ag} = r_a/r_g$, the $FE1 = e^{\frac{\left(1+\frac{1}{{\lambda_{ag}}^2}\right)^{\frac{-1}{2}}}{(1-\phi_w-\phi_a-\phi_b)^\beta}}$ and $FE2 = e^{\frac{\left(1+\left(\frac{\lambda_{ab}}{\lambda_{ag}}\right)^2\right)^{\frac{-1}{2}}}{(1-\phi_w-\phi_a-\phi_b)^\beta}}$, and $f(\phi_w, \phi_a, \phi_b, \phi_g)$ is the free energy density model. Therefore, the fluxes of water $J_w = \phi_w V_w$, the small $J_a = \phi_a V_a$ and the large osmolytes $J_b = \phi_b V_b$ in Eq. (4) in the main text are obtained according to Eqs. (S4)-(S16).

In the velocity expressions Eqs. (S5) and (S6), the terms $(d_{111} df_{wa} + d_{112} df_{aa} + d_{113} df_{ab}) * \frac{\partial \phi_a}{\partial r}$ and $(d_{221} df_{wb} + d_{222} df_{ab} + d_{223} df_{bb}) * \frac{\partial \phi_b}{\partial r}$ come from the self-interaction (virial coefficients $a_{aa}$ and $a_{bb}$ in Eq. (3) in the main text), and the terms $(d_{111} df_{wb} + d_{112} df_{ab} + d_{113} df_{bb}) * \frac{\partial \phi_b}{\partial r}$ and $(d_{221} df_{wa} + d_{222} df_{aa} + d_{223} df_{ab}) * \frac{\partial \phi_a}{\partial r}$ arise from the cross interaction (virial coefficients $a_{ab} = a_{ba}$ in Eq. (3) in the main text).

## II: Free energy density model

As stated in the main text, the free energy density can be simply written in the form for the colloidal gel materials [4,5]

$$f = k_B T \left( \sum_{i=w,a,b} \frac{1}{\sigma_i} \phi_i \ln \phi_i + \sum_{\substack{i=a,b,\\ j=a,b}} \frac{1}{\sigma_i \sigma_j} a_{ij} \phi_i \phi_j + G_0 \left( \frac{\phi_g - \phi_{g0}}{\phi_{g0}} \right)^2 \right), \tag{S17}$$

where $\sigma_i$ are the individual volumes of the components, $G_0 = m k_B T / \sigma_g$ is the elastic modulus with $m$ crosslinking degree.

The free energy density model can be rewritten as $f = \frac{k_B T}{\sigma_w} \tilde{f}$ with $\tilde{f}$ being the dimensionless form

$$\tilde{f} = \frac{1}{{\lambda_{aw}}^3} \left[ {\lambda_{aw}}^3 \phi_w \ln \phi_w + \phi_a \ln \phi_a + {\lambda_{ab}}^3 \phi_b \ln \phi_b + 4{\phi_a}^2 + 4{\lambda_{ab}}^3 {\phi_b}^2 \right.$$
$$+ {\lambda_{ab}}^3 \left(1 + \frac{1}{\lambda_{ab}}\right)^3 \phi_a \phi_b$$
$$\left. + \alpha \left( \frac{\phi_w + \phi_a + \phi_b - \phi_{w0} - \phi_{a0} - \phi_{b0}}{1 - \phi_{w0} - \phi_{a0} - \phi_{b0}} \right)^2 \right], \quad (S18)$$

where the dimensionless elasticity factor is $\alpha = m{\lambda_{ag}}^3$. Note that, since the small osmolyte radius $r_a$ =3 nm and the gyration radius of a segment $r_g$ =30 nm are both fixed, the ratio $\lambda_{ag} = r_a/r_g$ is also fixed in our problem, and then the mechanical elasticities in the nucleoplasm and the cytoplasm can be altered by tuning the value of crosslinking degree $m$. Besides, the free energy density model with the present form can be valid for the small elastic deformations of the gel-network. Usually, the cell with the cytoskeletons inside could maintain the small deformations in its volume under external osmotic shocks. Then, the free energy density model currently used is fully available for our problem.

**III: Soft-Cell approach (SCA) with dimensionless procedure**

SCA is developed on the basis of Lagrangian framework, and is capable of addressing the gel dynamics involving the coexistence of multi-components diffusions and elastic deformation of gel network, as termed diffusio-mechanical coupling (DMC) regime [6]. One outstanding advantage of employing SCA is that the resulting moving interface of the gel materials can be directly determined via the conservation law of the crosslinked polymers (or non-osmotic phases) retained in the gels instead of solving additional boundary condition which is often derived with a complex function of the multi-components' concentrations. In short, the employment of SCA not only enables a dynamic model to read elegantly, but also ensures the convenience in performing the numerical procedures. Such an approach has been invoked to address the dynamic problems for colloidal gel materials [5] and also microgels [2] in our previous work.

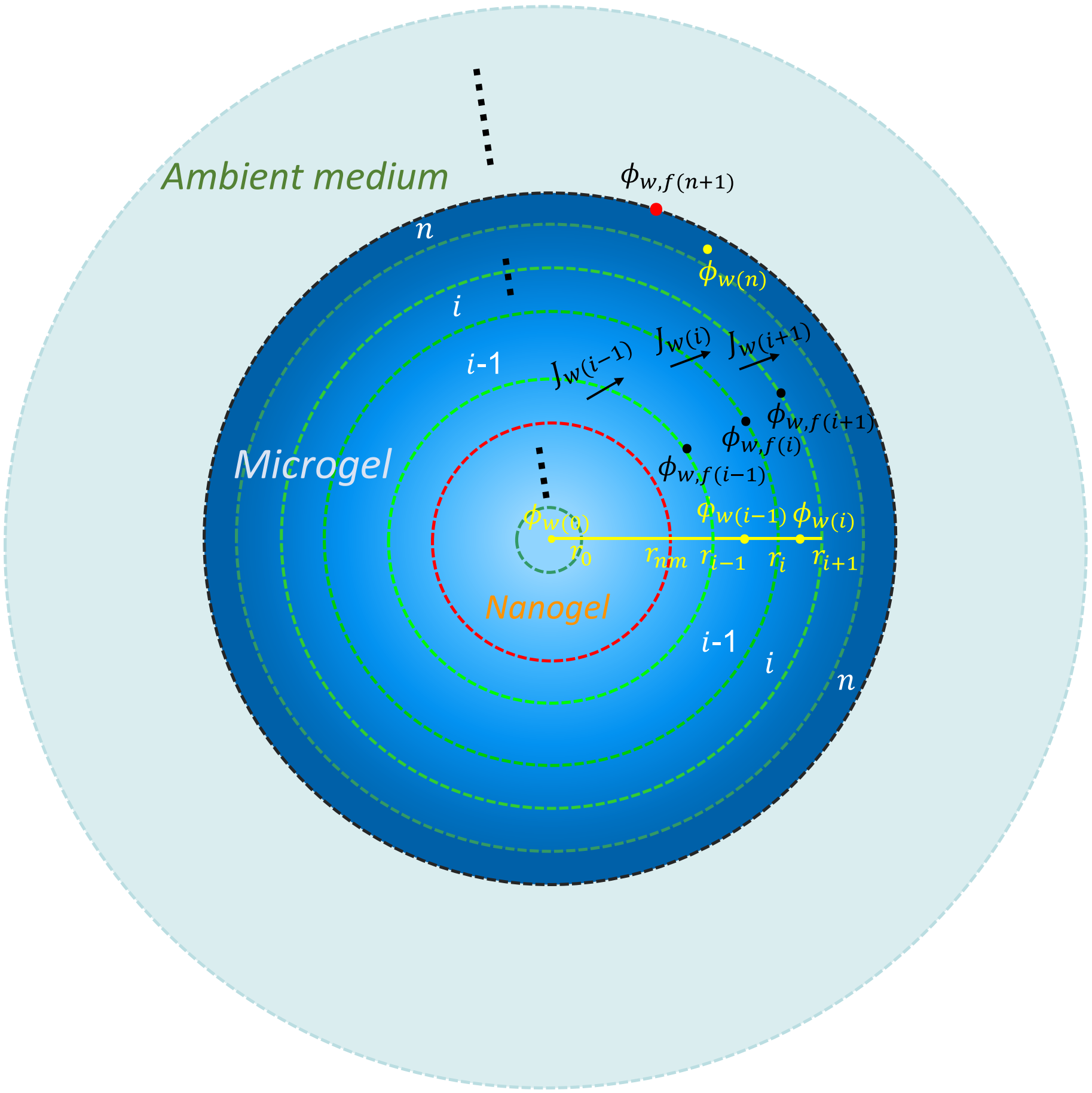

Fig. S1. Schematic illustration of SCA with taking the diffusive flux of water component for example (the small and the large osmolytes components follow the same representations). Water $J_{w(i)}$, as well as the small and the large osmolytes, diffuses either towards or outwards the nanogel and the microgel under osmotic shocks by crossing the crosslinked gel network. The diffusive flux relative to the crosslinked gel network has been guided by black arrow. $\phi_{w(i)}$ and $\phi_{w,f(i)}$ are the concentrations at the master node in the center of cell $i$ and at the face of cell $i$, respectively.

The SCA is implemented by conducting the dimensionless process for the governing equations (4) in the main text by invoking the simple variables as following

$$\tilde{r} = \frac{r}{R_{C0}}, \qquad c\tau = t\frac{D_{aw}}{{R_{C0}}^2}, \tag{S19}$$

where $R_{C0}$ is the initial radius of cell in the isotonic medium, and the dimensionless variable $\tilde{r}$ inside the cell meets $0 \le \tilde{r} \le 1$, and it inside the nucleus meets $0 \le \tilde{r} \le R_{N0}/R_{C0}$ with $R_{N0}$ being the initial radius of the nucleus in the isotonic medium, and $c$ is a factor to match real world time (here, we only focus on the case that all the time-dependent variables are discussed in the dimensionless time domains $\tau$, and then we set $c = 1$).

We consider a cell with an initial size $R_{C0}$ =3 μm, i.e., its volume $\sigma_{C0}$ ≈113 μm$^3$ [7], for the most of calculations in the main text. In order to avoid ambiguity in representing the SCA procedure, we dictate that the cell and the nucleus are replaced by the concepts of the microgel and the nanogel, respectively, as shown in Fig. 1 in the main text, and the word "cell" appearing in this section signifies a computing unite involved in the SCA procedure.

Consider 1-D spherical symmetry, as shown in Fig. S2, we first divide the domain ranging from the nanogel enter to the microgel border (or regarded as inner side of cell membrane) into $n$ cells with the uniform size, i.e., the concentric circular shells with a width per shell $\Delta\tilde{r}_{0(i)} = 1/n$ (the gel domain has been scaled by the initial radius $R_{C0}$ of the microgel). The initial border of the nanogel is defined as $\tilde{r}_{0(nm)} = R_{N0}/R_{C0}$. Fluxes of water, the small and the large osmolytes proceed crossing every cell as the microgel either shrinks or swells, while the gel concentration occupying in each cell is conserved. As a result, the width of the cell $i$ at the current state of time $\tau$ is updated by $\Delta\tilde{r}_i$.

As seen in Fig. S2 for the cross-section of the concentric circular shells, the current cell $i$ has two faces located by point $\tilde{r}_i$ (left) and $\tilde{r}_{i+1}$ (right), and also contains two sorts of concentrations: the one at the master node $\phi_{w(i)}$ which occupies in the cell $i$, and another one locates at the configured nodes, such as $\phi_{w,f(i)}$ at the left face and $\phi_{w,f(i+1)}$ at the right face (the same configurations are applied to the osmolytes). Based on the derivations in section I, the dimensionless fluxes of water, the small and the large osmolytes across the current cell face $i$ are written into the common formats

$$\tilde{J}_{w(i)} = \phi_{w,f(i)}\left(\widetilde{D}_{w1} * \left.\frac{\partial\phi_w}{\partial\tilde{r}}\right|_i + \widetilde{D}_{w2} * \left.\frac{\partial\phi_a}{\partial\tilde{r}}\right|_i + \widetilde{D}_{w3} * \left.\frac{\partial\phi_b}{\partial\tilde{r}}\right|_i\right), \tag{S20}$$

$$\tilde{J}_{a(i)} = \phi_{a,f(i)}\left(\widetilde{D}_{a1} * \left.\frac{\partial\phi_w}{\partial\tilde{r}}\right|_i + \widetilde{D}_{a2} * \left.\frac{\partial\phi_a}{\partial\tilde{r}}\right|_i + \widetilde{D}_{a3} * \left.\frac{\partial\phi_b}{\partial\tilde{r}}\right|_i\right), \tag{S21}$$

$$\tilde{J}_{b(i)} = \phi_{b,f(i)}\left(\widetilde{D}_{b1} * \left.\frac{\partial\phi_w}{\partial\tilde{r}}\right|_i + \widetilde{D}_{b2} * \left.\frac{\partial\phi_a}{\partial\tilde{r}}\right|_i + \widetilde{D}_{b3} * \left.\frac{\partial\phi_b}{\partial\tilde{r}}\right|_i\right), \tag{S22}$$

where the dimensionless fluxes are obtained by $\tilde{J}_{w(i)} = \frac{R_{C0}}{D_{aw}} J_{w(i)}$, $\tilde{J}_{a(i)} = \frac{R_{C0}}{D_{aw}} J_{a(i)}$, and $\tilde{J}_{b(i)} = \frac{R_{C0}}{D_{aw}} J_{b(i)}$, and the dimensionless diffusivity candidates $\widetilde{D}_{ij}$ $(i = w, a, b; j = 1,2,3)$ can be readily obtained from the expressions of the velocities represented in section I.

The concentrations of water, the small and the large osmolytes involved in the $\widetilde{D}_{ij}$ in Eqs. (S20)-(S22) are used by the value at the face $i$, i.e., $\phi_{w,f(i)}$, $\phi_{a,f(i)}$ and $\phi_{b,f(i)}$. These concentrations at the face $i$ can be approximately determined by the linear interpolation of the concertation at the master nodes of the neighboring cells $i-1$ and $i$, yielding

$$\phi_{w,f(i)} = \frac{\Delta\tilde{r}_{i-1}\phi_{w(i)} + \Delta\tilde{r}_i\phi_{w(i-1)}}{\Delta\tilde{r}_{i-1} + \Delta\tilde{r}_i}, \tag{S23}$$

$$\phi_{a,f(i)} = \frac{\Delta\tilde{r}_{i-1}\phi_{a(i)} + \Delta\tilde{r}_i\phi_{a(i-1)}}{\Delta\tilde{r}_{i-1} + \Delta\tilde{r}_i}, \tag{S24}$$

$$\phi_{b,f(i)} = \frac{\Delta\tilde{r}_{i-1}\phi_{b(i)} + \Delta\tilde{r}_i\phi_{b(i-1)}}{\Delta\tilde{r}_{i-1} + \Delta\tilde{r}_i}, \tag{S25}$$

Moreover, the concentration gradients over the face $i$ in Eqs. (S20)-(S22) can be given by the simple differential rule

$$\left.\frac{\partial\phi_w}{\partial\tilde{r}}\right|_i = \frac{\phi_{w(i)} - \phi_{w(i-1)}}{(\Delta\tilde{r}_i + \Delta\tilde{r}_{i-1})/2}, \tag{S26}$$

$$\left.\frac{\partial\phi_a}{\partial\tilde{r}}\right|_i = \frac{\phi_{a(i)} - \phi_{a(i-1)}}{(\Delta\tilde{r}_i + \Delta\tilde{r}_{i-1})/2}, \tag{S27}$$

$$\left.\frac{\partial\phi_b}{\partial\tilde{r}}\right|_i = \frac{\phi_{b(i)} - \phi_{b(i-1)}}{(\Delta\tilde{r}_i + \Delta\tilde{r}_{i-1})/2}. \tag{S28}$$

We can subsequently solve the time evolution equations of water, the small and the large osmolytes with respect to the current cell $i$ at the current state $\tau_+ = \tau + \Delta\tau$ with the time interval $\Delta\tau$

$$\phi_{w(i)}(\tau_+) = \phi_{w(i)}(\tau) + 3\Delta\tau\frac{\tilde{J}_{w(i)}\tilde{r}_i^{\,2}(\tau_+) - \tilde{J}_{w(i+1)}\tilde{r}_{i+1}^{\,2}(\tau_+)}{\tilde{r}_{i+1}^{\,3}(\tau_+) - \tilde{r}_i^{\,3}(\tau_+)}, \tag{S29}$$

$$\phi_{a(i)}(\tau_+) = \phi_{a(i)}(\tau) + 3\Delta\tau\frac{\tilde{J}_{a(i)}\tilde{r}_i^{\,2}(\tau_+) - \tilde{J}_{a(i+1)}\tilde{r}_{i+1}^{\,2}(\tau_+)}{\tilde{r}_{i+1}^{\,3}(\tau_+) - \tilde{r}_i^{\,3}(\tau_+)}, \tag{S30}$$

$$\phi_{b(i)}(\tau_+) = \phi_{b(i)}(\tau) + 3\Delta\tau\frac{\tilde{J}_{b(i)}\tilde{r}_i^{\,2}(\tau_+) - \tilde{J}_{b(i+1)}\tilde{r}_{i+1}^{\,2}(\tau_+)}{\tilde{r}_{i+1}^{\,3}(\tau_+) - \tilde{r}_i^{\,3}(\tau_+)}. \tag{S31}$$

Herein, as the microgel, as well as the nanogel, changes the volume by adapting to the osmotic shocks, the width $\Delta\tilde{r}_i = \tilde{r}_{i+1} - \tilde{r}_i$ of the cell $i$ must alter over time, meaning the "Soft-Cell" nature, which differs from the conventional finite volume method. The variation in size of the current cell is determined by the conservation law of the crosslinked polymers (gel) retained in the cell $\Delta\tilde{r}_i$ with respect to that in the initial cell $\Delta\tilde{r}_{0(i)}$

$$\left(\tilde{r}_{i+1}^{\,3} - \tilde{r}_i^{\,3}\right)\phi_{g(i)}(\tau_+) = \left(\tilde{r}_{0(i+1)}^{\,3} - \tilde{r}_{0(i)}^{\,3}\right)\phi_{g0}, \tag{S32}$$

where $\phi_{g(i)}(\tau_+) = 1 - \phi_{w(i)}(\tau_+) - \phi_{a(i)}(\tau_+) - \phi_{b(i)}(\tau_+)$ and $\phi_{g0}$ are the gel concentrations at the current state and the initial state, respectively. Note that, all the variables at right side of Eq. (S32) are the constants initially assigned, and particularly the first cell ($i = 1$) has been regarded as a complete sphere instead of the spherical shell ($i > 1$), hence the conservation law applied to this cell ($i = 1$) reads $\tilde{r}_1{}^3\phi_{g(1)} = r_{0(1)}{}^3\phi_{g0}$ with $\phi_{g(1)} = 1 - \phi_{w(1)} - \phi_{a(1)} - \phi_{b(1)}$.

The dimensionless boundary conditions corresponding to the original ones stated in the main text for our problems are listed as following:

(i): At the nanogel center, i.e., at the first cell $\tilde{r}_1$, $\left.\frac{\partial\phi_w}{\partial\tilde{r}}\right|_{i=1} = \left.\frac{\partial\phi_a}{\partial\tilde{r}}\right|_{i=1} = \left.\frac{\partial\phi_b}{\partial\tilde{r}}\right|_{i=1} = 0$, meaning the zero fluxes.

(ii): At the CM, i.e., at the border of the microgel $\tilde{r}_{n+1}(\tau)$, $\tilde{J}_{a(\tilde{r}_{n+1})} = \tilde{J}_{b(\tilde{r}_{n+1})} = 0$, and $\tilde{J}_{w(\tilde{r}_{n+1})} = -\,d\tilde{R}_C(\tau)/d\tau$, where $\tilde{R}_C(\tau) = R_C(\tau)/R_{C0} = exp\big(Cont\ \Delta\tilde{\Pi}\ \tau\big)$, as mentioned in the main text.

(iii): At the NE, i.e., at the border of the nanogel $\tilde{r}_{nm}(\tau)$, the diffusions of the large osmolytes between the cytoplasm and the nucleoplasm are not allowed, i.e., $\tilde{J}_{b(\tilde{r}_{nm})} = 0$.

(iv): At the NE $\tilde{r}_{nm}(\tau)$, we let $\left.\frac{\partial\phi_b}{\partial\tilde{r}}\right|_{i=\tilde{r}_{nm}} = 0$ in the fluxes of water and the small osmolytes, i.e., Eqs. (S20) and (S21).

Performing the time iteration calculations using above set of equations with the appropriate boundary conditions discussed above, we can produce all needed spatiotemporal variables, such as $\phi_w(\tau,\tilde{r})$, $\phi_a(\tau,\tilde{r})$, $\phi_b(\tau,\tilde{r})$, $\tilde{r}_{nm}(\tau) = R_N(\tau)/R_{C0}$ and $\tilde{r}_{n+1}(\tau) = R_C(\tau)/R_{C0}$ simultaneously. As discussed earlier, the moving conditions of the microgel border $\dot{R}_C(\tau)$ and the nanogel border $\dot{R}_N(\tau)$ can be directly accessed by $d\tilde{r}_{n+1}(\tau)/d\tau$ and $d\tilde{r}_{nm}(\tau)/d\tau$, respectively, in which the elastic deformable behaviors of the bulk microgel and multi-diffusions behaviors are both involved.

## IV: Some supplementary results

### 1. The effects of initial N/C ratio $\varepsilon_0$

We probe here if the initial N/C ratio, defined by $\varepsilon_0 = R_{N0}/R_{C0}$ in the main text, has the influence on the dynamics of the N/C ratio. Figure S2 shows the effect of $\varepsilon_0$ upon the N/C ratio dynamics for the condition of the equal initial osmolytes concentrations $\phi_{a0} = \phi_{b0}$ =0.13. Apparently, for different $\varepsilon_0$, the dynamic profile of the N/C ratio behaves rather similarly, i.e., the nonmonotonic variations with time under the osmotic shocks. Besides, the variable $\varepsilon_0$ cannot alter the maintenance behavior of the N/C ratio between the value at the initial state and the steady state, as discussed in the matin text, suggesting that such a N/C ratio maintenance is exclusively determined by the initial concentrations of the large osmolytes localized to the nucleoplasm and the cytoplasm. The phenomenon presented here that the initial N/C ratio has the little effects on the dynamic behaviors (or equilibrium behaviors) of the N/C ratio can be supported by the experimental measurements in Ref. [7] (e.g., the Fig. 7 in Ref. [7]).

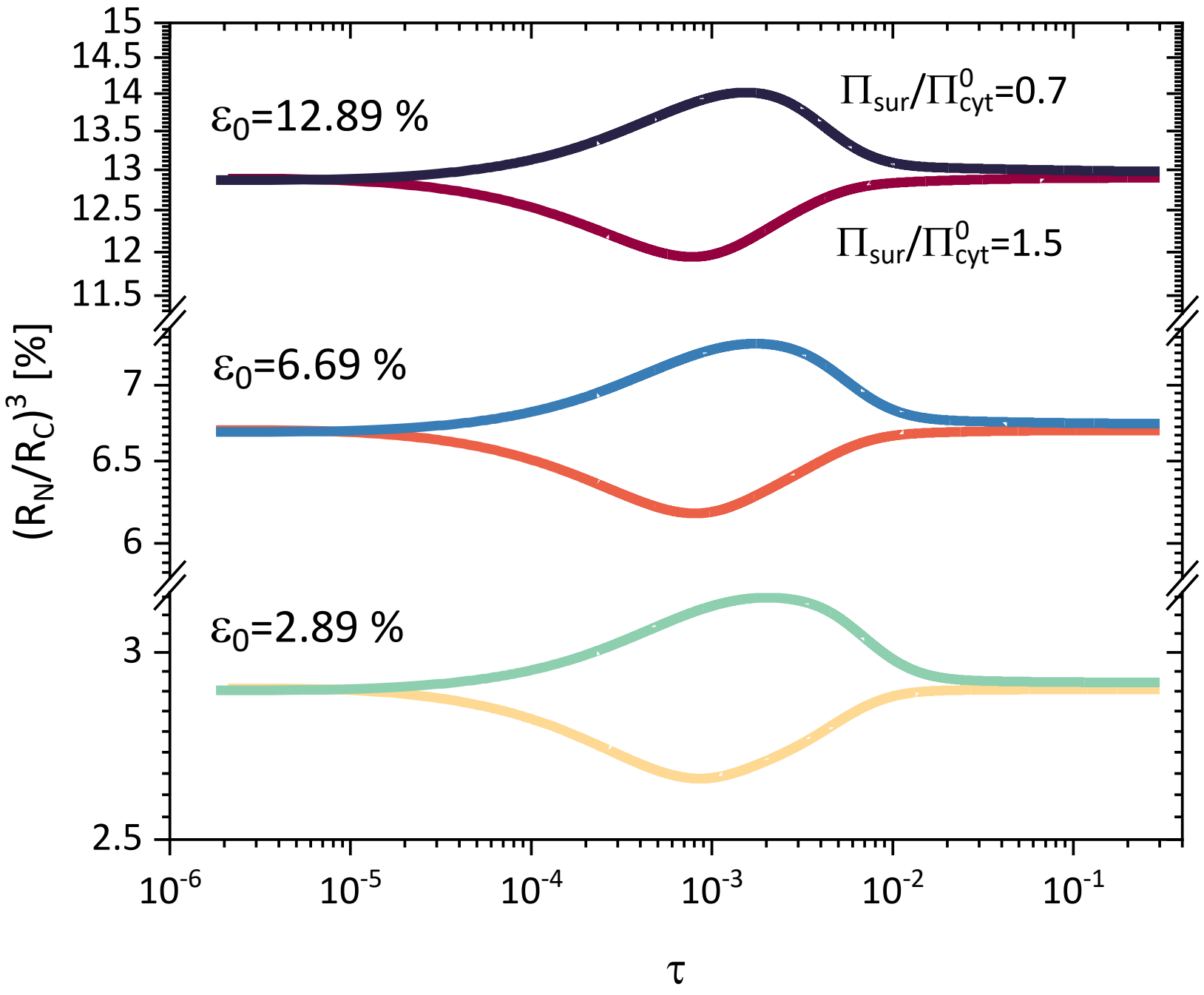

FIG. S2. Dynamic profiles of the N/C ratio under the hyperosmotic and the hypoosmotic shocks for different initial N/C ratios $\varepsilon_0$. Parameters are $\phi_{w0}$ =0.7, $\phi_{a0} = \phi_{b0}$ =0.13, $r_a$ =3 nm, $r_b$ =20 nm.

## 2. The more general case $\boldsymbol{\phi_{b0}^{nuc} \neq \phi_{b0}^{cyt}}$ ($\boldsymbol{\bar{\phi}_b^{nuc} \neq \bar{\phi}_b^{cyt}}$ at the steady state)

In the main text, we have discussed the special case in which the large osmolytes separated into the cytoplasm and the nucleoplasm by the nucleus envelope have identical initial concentrations $\phi_{b0}^{nuc} = \phi_{b0}^{cyt} = 0.13$ ($\phi_{w0}$ =0.7 and $\phi_{a0}$ =0.13), also indicating the observed comparable average concentrations of the large osmolytes at the steady state $\bar{\phi}_b^{nuc} = \bar{\phi}_b^{cyt}$. In order to ensure repeatability of the relationship between the large osmolytes concentrations $\bar{\phi}_b^{nuc}$ and $\bar{\phi}_b^{cyt}$ at the steady state for this special case, we represent it here by considering another condition of the identical initial concentrations of the large osmolytes: $\phi_{b0}^{nuc} = \phi_{b0}^{cyt} = 0.1$ ($\phi_{w0}$ =0.7 and $\phi_{a0}$ =0.15). Figure S3 (a) shows the dynamic profiles of the N/C ratio for different osmotic shocks. Clearly, the dynamic behaviors of the N/C ratio observed here are identical to those represented in Fig. 2 (c) in the main text, i.e., the nonmonotonic variations in the N/C ratio with time and also the maintenance of the N/C ratio at the steady state to its initial value are both reproduced.

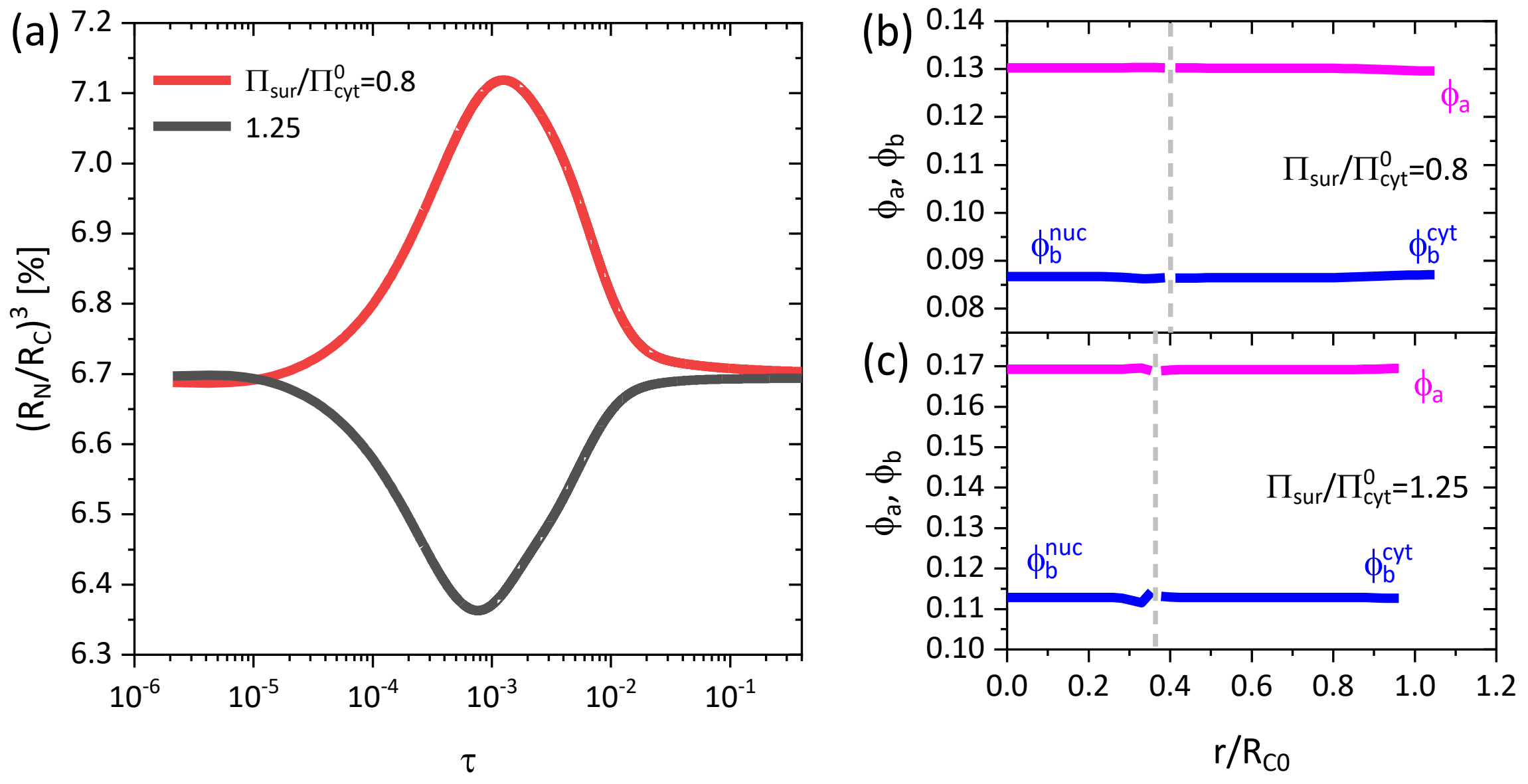

FIG. S3. (a) Dynamics of the N/C ratio for the special case, i.e., the identical initial concentrations of the large osmolytes, under the osmotic shocks. Parameters used here are: $\phi_{b0}^{cyt} = \phi_{b0}^{nuc}$ =0.1, $\phi_{w0}$ =0.7, $\phi_{a0}$ =0.15, $r_a$ =3 nm, $r_b$ =20 nm and $R_{N0}/R_{C0}$ =0.406. Concentration profiles of the small and the large osmolytes at the steady state ($\tau$ =0.3) for (b) the hypoosmotic shock and (c) the hyperosmotic shock. The dashed lines in (b) and (c) guide the locations of the nucleus envelope $R_N/R_{C0}$, and the domain on the left side of the dashed line represents the nucleoplasm and that on the right side of the dashed line represents the cytoplasm.

As discussed in the main text, the N/C ratio can be represented by $\varepsilon(t) = \varepsilon_0/(\varepsilon_0(1-k)+k)$, where $\varepsilon_0 = R_{N0}/R_{C0}$ is an initial N/C ratio and $k = \phi_{b0}^{cyt}\bar{\phi}_b^{nuc}/\phi_{b0}^{nuc}\bar{\phi}_b^{cyt}$, where $\bar{\phi}_b^{nuc}(t) = 3\int_0^{R_N} r^2\phi_b^{nuc}dr/{R_N}^3$ and $\bar{\phi}_b^{cyt}(t) = 3\int_{R_N}^{R_C} r^2\phi_b^{cyt}dr/\left({R_C}^3 - {R_N}^3\right)$ are assumed to be the volumetric average concentrations with spatial independence. As seen in Figs. S3 (b) and S3 (c), these average concentrations of the large osmolytes $\bar{\phi}_b^{nuc}$ and $\bar{\phi}_b^{cyt}$ at the steady state ($\tau$ =0.3) are rather comparable, giving the ratio $k$ =1. Therefore, $\varepsilon(\tau = 0.3) = \varepsilon_0$, i.e., the N/C ratio at the steady state maintains the same value as the initial one under external osmotic shocks for this special case.

Here, we did the calculations for the more general case of unequal initial concentrations of the large osmolytes $\phi_{b0}^{nuc} \neq \phi_{b0}^{cyt}$, and also show how the dynamics of the N/C ratio behaves for this case. Generally, the concentration of the macromolecules in the nucleoplasm could be slightly higher than that in the cytoplasm, and we then set here $\phi_{b0}^{nuc}$ =0.14 and $\phi_{b0}^{cyt}$ =0.1 ($\phi_{w0}$ =0.7 and $\phi_{a0}$ =0.12). As shown in Fig. S4 (a), although the nonmonotonic behaviors of the N/C ratio dynamics have been remained, the values of the N/C ratio at the steady state are seen to evidently deviate from those at the initial state, suggesting that the maintenance of the N/C ratio for the special case has been broken. Figures S4 (b) and S4 (c) show the concentrations of the small and the large osmolytes at the steady state for the general case.

By revisiting the formular of $\varepsilon(t)$ mentioned above, according to the concentration profiles in Figs. S4 (b) and S4 (c), we can obtain $k$ =1.072 for the hypoosmotic shock and $k$ =1.089 for the hyperosmotic shock, and then the N/C ratio at the steady state can be approximately estimated by $\varepsilon(t \to \infty) \cong \varepsilon_0/k$, i.e., we obtain $\varepsilon(t \to \infty) \approx$6.24 % for the hypoosmotic shock and $\varepsilon(t \to \infty) \approx$6.14 % for the hyperosmotic shock, which are relatively close to the numerical measurements as shown in Fig. S4 (a).

Alternatively, as suggested in the main text, the N/C ratio for this general case can be readily estimated by

$$\varepsilon(t) \cong \varepsilon_0 \frac{\phi_{b0}^{cyt}}{\phi_{a0}} \frac{\bar{\phi}_a(t)}{\bar{\phi}_b^{cyt}(t)}. \tag{S33}$$

To produce Eq. (S33), we have used the conservation law of the small osmolytes within the whole cell domain ${R_C}^3 = {R_{C0}}^3 \, \phi_{a0}/\bar{\phi}_a(t)$, where $\bar{\phi}_a(t) = 3\int_0^{R_C} r^2 \phi_a(t) dr/{R_C}^3$. In Eq. (S33), we can re-define the factor $k = \phi_{b0}^{cyt}\, \bar{\phi}_a(t)/\phi_{a0}\bar{\phi}_b^{cyt}(t)$. Again, according to the concentration profiles of the osmolytes in Figs. S4 (b) and S4 (c), the N/C ratio at the steady state, using Eq. (S33), can be evaluated as $\varepsilon(t \to \infty) \approx$6.26 % ($k \approx$0.936) for the hypoosmotic shock and $\varepsilon(t \to \infty) \approx$6.33 % ($k \approx$0.947) for the hyperosmotic shock. Eq. (S33) suggests that the N/C ratio, especially at the steady state, can be evaluated well by only accessing the information, such as the average concentrations of the osmolytes, in the cytoplasm instead of the nucleoplasm, which is experimentally favorable. In summary, our results show that if the concentrations of the large osmolytes (i.e., the bio-macromolecules) preferentially localized to the nucleoplasm and the cytoplasm are not comparable initially, their concentrations at the steady state at which the cell has adapted to the osmotic shocks still appear with a significant difference across the nucleus envelope. This more general case discussed here can not only make the dynamics of the N/C ratio more complex, but also disrupt the maintenance of the N/C ratio at between the initial state and the steady state. We speculate that the reason why only the factors, such as the concentration and the size of the osmolytes (Fig. 3 (d) in the main text), give rise to these complicated dynamic (or the equilibrium) behaviors of the N/C ratio is that these factors appearing as the ingredients in the free energy density model can modulate directly the interactions between intracellular molecules (bio-osmolytes). The further explorations for clarifying this complicated case will appear on the next work.

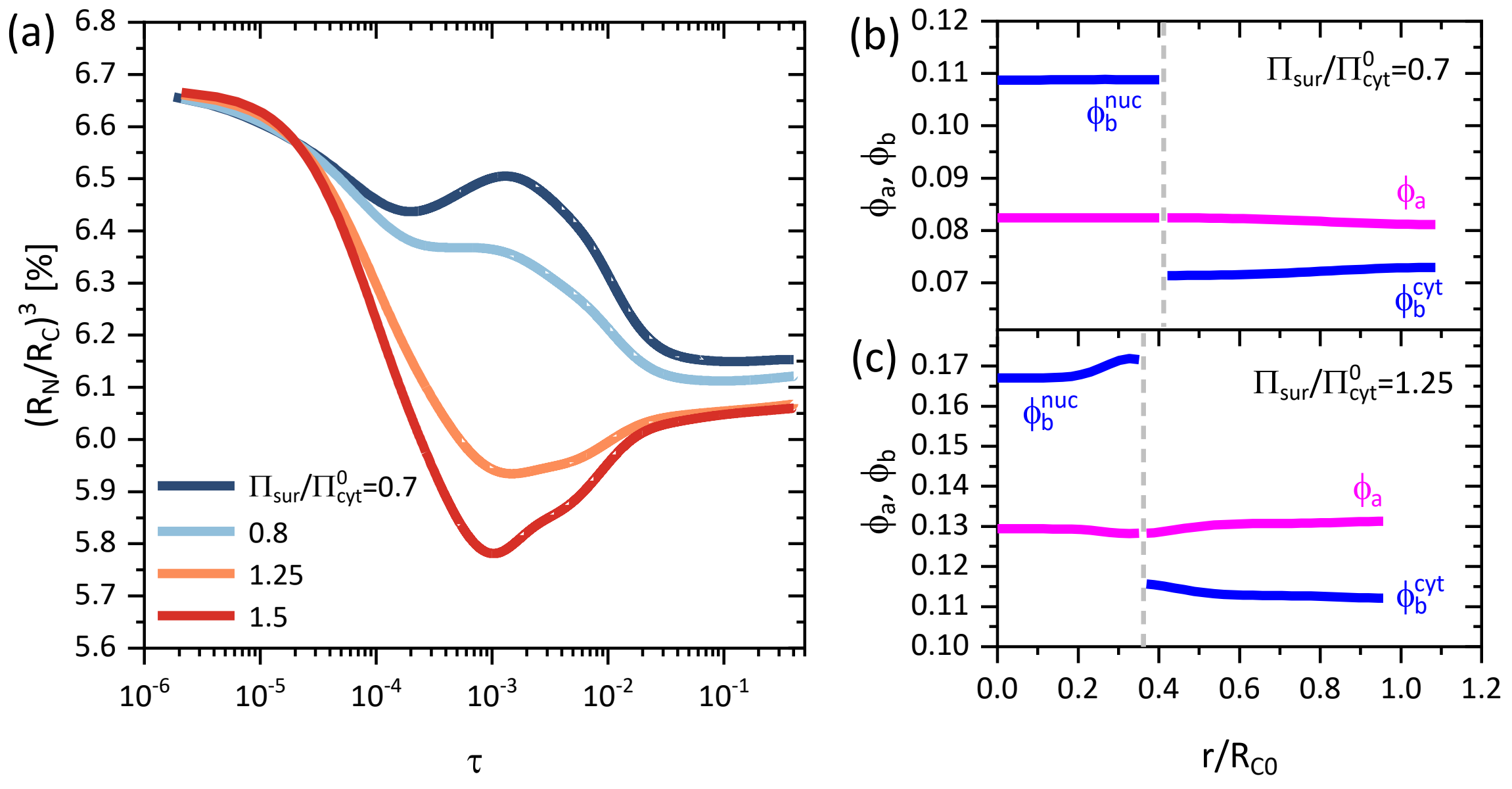

FIG. S4. (a) Dynamics of the N/C ratio for the more general case, i.e., the unequal initial concentrations of the large osmolytes, under the osmotic shocks. Parameters used here are: $\phi_{b0}^{cyt} = 0.1$, $\phi_{b0}^{nuc} = 0.14$, $\phi_{w0} = 0.7$, $\phi_{a0} = 0.12$, $r_a = 3$ nm, $r_b = 20$ nm and $R_{N0}/R_{C0}$ =0.406. Concentration profiles of the small and the large osmolytes at the steady state ($\tau$ =0.3) for (b) the hypoosmotic shock and (c) the hyperosmotic shock. The dashed lines in (b) and (c) guide the locations of the nucleus envelope $R_N/R_{C0}$, and the domain on the left side of the dashed line represents the nucleoplasm and that on the right side of the dashed line represents the cytoplasm.

## References

[1] T. Bertrand, J. Peixinho, S. Mukhopadhyay, and C. W. MacMinn, Phy. Rev. Applied **6**, 064010 (2016).

[2] J. Z. Sui, Phys. Chem. Chem. Phys. **25**, 410 (2023).

[3] T. Kalwarczyk, N. Ziębacz, A. Bielejewska, E. Zaboklicka, K. Koynov, J. Szymański, A. Wilk, A. Patkowski, J. Gapiński, H.-J. Butt and R. Hołyst, Nano Lett. **11**, 2157 (2011).

[4] J.-P. Hansen, and I. R. McDonald, Theory of simple. liquids with applications to soft matter (Fourth Edition), Academic Press of Elsevier, Oxford, UK, 2013.

[5] J. Z. Sui, Phys. Chem. Chem. Phys. **22**, 14340 (2020).

[6] M. Doi, Soft Matter Physics, Oxford University Press, Oxford, 2013.

[7] J. Lemière, P. Real-Calderon, L. J. Holt, T. G. Fai, and F. Chang, eLife **11**, E76075 (2022).